\documentclass[12pt,aps,nofootinbib]{revtex4}

\usepackage{amsmath,amssymb,bm}
\usepackage[dvips]{graphicx}
\usepackage{hyperref}
\usepackage{yfonts}
\usepackage{color}

\newcommand{\beq}{\begin{eqnarray}}
\newcommand{\eeq}{\end{eqnarray}}
\renewcommand\d{\partial}

\begin{document}

\title{Chirality Driven Helical Pattern Formation}

\author{Naoki Yamamoto}
\affiliation{Department of Physics, Keio University, Yokohama 223-8522, Japan}
\begin{abstract}
We study the pattern formation of chiral charges in the presence of reactions. We show that, in 
contrast to the original Turing's mechanism of pattern formation in diffusion-reaction systems, 
the interplay between chiral effects and reactions can lead to a new kind of instability against 
spatially inhomogeneous perturbations, and furthermore, to a helical pattern formation, even 
without diffusion. This provides a new physical mechanism that can generate a macroscopic 
helical structure from microscopic chirality, including chirality of elementary particles, via 
nonequilibrium processes.
\end{abstract}
\maketitle

\section{Introduction}
The notion of chirality is universally important in various areas of natural sciences ranging from 
physics and chemistry to biology. The helical motions and helical structures that possess right- or 
left-handed chirality appear over hierarchical scales from elementary particles (e.g., neutrinos), 
chemical molecules (e.g., amino acids) and biological polymers (e.g., DNA) to biological 
architectures (e.g., shells). It is generally considered that macroscopic helical structures are 
tied to microscopic chirality of the constituents. Although theoretical and experimental studies 
have been made on such connections for specific systems mostly in chemistry and biology, 
the generic physical mechanism to bridge between the different hierarchies has been elusive. 
In particular, the possible emergence of macroscopic helical structures from the most microscopic 
chirality to date---chirality of elementary particles---has been poorly understood.%
\footnote{For several hypotheses on the possible effects of the parity violation by the weak 
interaction on macroscopic chemical and biological helical structures, see, e.g., 
Refs.~\cite{Bonner2000, Avalos2000}.}

In this paper, we provide a fundamentally new physical mechanism for the emergence of the 
macroscopic helical structure from microscopic chirality, including chirality of elementary particles, 
via nonequilibrium processes. We show that the chiral charges, in the presence of reactions 
and under certain conditions, lead to a new type of instability against spatially inhomogeneous 
perturbations, and furthermore, to a \emph{helical} pattern formation, \emph{even without} 
diffusion. Our mechanism should be contrasted with the Turing's mechanism of pattern 
formation \cite{Turing, Murray} where diffusion plays an essential role. Our analysis is based on 
the effective theory and is independent of the details of systems. Hence, it is applicable to 
generic systems involving chiral charges, as long as the conditions that we shall derive are satisfied.

At a more technical level, the question that we address here can be seen from a different 
viewpoint. The conventional Turing instability and pattern formation \cite{Turing, Murray} is 
based on the diffusion equation that follows from the Fick's law of diffusion, 
${\bm j} = - D{\bm \nabla} n$ with $D$ the diffusion constant and $n$ the charge density, which 
necessarily generates entropy. However, it is known that, in some cases, there appear 
dissipationless currents originating from some topological nature of a system (see below). 
Then, one may ask a possible pattern formation due to such topological currents instead of 
diffusion currents. As we will see, our results provide an answer to such a question for the 
specific case where topological currents occur due to the chirality of elementary particles.

\section{Effective theory for chiral charges with reactions}
To illustrate the essence of our mechanism for helical pattern formation, let us consider a 
1+1 dimensional system. It is straightforward to extend our argument to other odd spatial 
dimensions (including 3+1 dimensions) where chirality is well defined. 

Let us consider an effective theory for generic right- and left-handed charges, 
$n_{\rm R}(x,t)$ and $n_{\rm L}(x,t)$, in an open system with reactions. 
We define the vector and axial charges, $n_{\rm V} \equiv n_{\rm R} + n_{\rm L}$ and 
$n_{\rm A} \equiv n_{\rm R} - n_{\rm L}$, which transform under the parity transformation as 
$n_{\rm V} \rightarrow n_{\rm V}$ and $n_{\rm A} \rightarrow -n_{\rm A}$, respectively.
In the presence of reactions, the generic effective theory for $n_{\rm V}$ and $n_{\rm A}$, 
which is consistent with parity symmetry, to leading order in derivatives is given by
\begin{subequations}
\label{EFT}
\begin{align}
\label{EFT1}
\d_t n_{\rm V} &= \alpha_1 \d_x n_{\rm A} 
+ \alpha_2 n_{\rm A} \d_x n_{\rm V} + \alpha_3 n_{\rm V} \d_x n_{\rm A}
+ f(n_{\rm V}, n_{\rm A}) + O(\d_x^2), \\
\label{EFT2}
\d_t n_{\rm A} &= \beta_1 \d_x n_{\rm V} 
+ \beta_2 n_{\rm V} \d_x n_{\rm V} + \beta_3 n_{\rm A} \d_x n_{\rm A}
+ g(n_{\rm V}, n_{\rm A}) + O(\d_x^2).
\end{align}
\end{subequations}
Here $\alpha_1$, $\alpha_2$, $\alpha_3$, $\beta_1$, $\beta_2$, and $\beta_3$ are some 
parity-invariant constants that depend on the microscopic details of the system, and 
$f(n_{\rm V}, n_{\rm A})$ and $g(n_{\rm V}, n_{\rm A})$ denote the reaction terms that are 
generically nonlinear functions of $n_{\rm V}$ and $n_{\rm A}$ and that make $n_{\rm V}$ 
and $n_{\rm A}$ nonconserved.%
\footnote{The $\alpha_{2,3}$ and $\beta_{2,3}$ terms are nonlinear functions of 
$n_{\rm V}$ and/or $n_{\rm A}$, but they are not included in the reaction terms. This is because 
they vanish in the limit of small momentum ${\bm q} \rightarrow {\bm 0}$ and they respect the 
charge conservation in the absence of $f$ and $g$.}
From the requirement of parity symmetry, $f$ and $g$ satisfy the conditions, 
$f(n_{\rm V}, -n_{\rm A}) = f(n_{\rm V}, n_{\rm A})$ and 
$g(n_{\rm V}, -n_{\rm A}) = -g(n_{\rm V}, n_{\rm A})$.
When some parity-breaking background field is present, additional terms, e.g., 
$\gamma_1 \d_x n_{\rm V}$ and $\gamma_2 \d_x n_{\rm A}$ with $\gamma_{1,2}$ being some 
parity-odd quantities, are added into the right-hand sides of Eqs.~(\ref{EFT1}) and (\ref{EFT2}), 
respectively. The microscopic origin of the $\alpha$ and $\beta$ terms and the detailed forms of 
$f(n_{\rm V}, n_{\rm A})$ and $g(n_{\rm V}, n_{\rm A})$ as well as those of parity-breaking 
background fields will be irrelevant to the following discussion.

Note that, in a system involving only vector-type charges, which is the usual situation 
considered in the context of pattern formation \cite{Turing, Murray}, the $\alpha$ and $\beta$ 
terms in Eqs.~(\ref{EFT}) would be absent; the presence of these terms are specific to the 
system with chirality, and hence, they will be called the chiral terms. 
Note also that the diffusion terms of the form $D \d_x^2 n_{\rm V, A}$, which are typically leading 
order in derivatives (except for reaction terms), are included in $O(\d^2)$ and are higher order 
compared with the chiral terms.

\section*{Example of chiral terms}
So far, our construction of the effective theory (\ref{EFT}) has been general. Before proceeding 
further, we discuss one concrete realization of the chiral terms above by the topological 
transport phenomena in relativistic matter of chiral fermions: the so-called the chiral magnetic 
effect (CME) \cite{Vilenkin:1980fu, Nielsen:1983rb, Fukushima:2008xe} and 
the chiral vortical effect (CVE) \cite{Vilenkin:1979ui, Son:2009tf, Landsteiner:2011cp}, 
which are the currents along the direction of an external magnetic field ${\bm B}$ and 
a vorticity ${\bm \omega}$, respectively. The generic expressions of the vector and axial 
currents due to the CME and CVE at finite vector and axial chemical potentials, 
$\mu_{\rm V}$ and $\mu_{\rm A}$, and at finite temperature $T$ are given by
\begin{align}
{\bm j}_{\rm V} &= \frac{\mu_{\rm A}}{2\pi^2}{\bm B + \frac{\mu_{\rm V} \mu_{\rm A}}{\pi^2}}{\bm \omega}\,, 
\\
\quad {\bm j}_{\rm A} &= \frac{\mu_{\rm V}}{2\pi^2}{\bm B} 
+ \left(\frac{\mu_{\rm V}^2 + \mu_{\rm A}^2}{2\pi^2} + \frac{T^2}{6} \right){\bm \omega}\,,
\end{align}
respectively. Owing to the topological nature of chiral fermions, the transport coefficients 
(except for the $T$-dependent term) are exact independently of interactions \cite{Son:2012wh}.
Also, these currents are \emph{dissipationless} and do not generate entropy \cite{Son:2009tf}.

Inserting these expressions into the continuity equations for $n_{\rm V}$ and $n_{\rm A}$,
\beq
\label{continuity}
\d_t n_{\rm V} + {\bm \nabla} \cdot {\bm j}_{\rm V} =0, \quad 
\d_t n_{\rm A} + {\bm \nabla} \cdot {\bm j}_{\rm A} =0, 
\eeq
we obtain
\begin{subequations}
\label{continuity_chiral}
\begin{align}
\d_t n_{\rm V} &= - \frac{1}{2\pi^2 \chi_{\rm A}} {\bm B} \cdot {\bm \nabla} n_{\rm A}
- \frac{1}{\pi^2 \chi_{\rm V} \chi_{\rm A}} 
(n_{\rm A} {\bm \omega} \cdot {\bm \nabla} n_{\rm V} + n_{\rm V} {\bm \omega} \cdot {\bm \nabla} n_{\rm A}) \,, 
\\
\d_t n_{\rm A} &= - \frac{1}{2\pi^2 \chi_{\rm V}} {\bm B} \cdot {\bm \nabla} n_{\rm V}
- \frac{1}{\pi^2 \chi_{\rm V}^2} n_{\rm V} {\bm \omega} \cdot {\bm \nabla} n_{\rm V} 
- \frac{1}{\pi^2 \chi_{\rm A}^2} n_{\rm A} {\bm \omega} \cdot {\bm \nabla} n_{\rm A}\,, \quad 
\end{align}
\end{subequations}
for a homogeneous temperature, where $\chi_{\rm V}$ and $\chi_{\rm A}$ are the 
susceptibilities defined by $\chi_{\rm V} \equiv {\d \mu_{\rm V}}/{\d n_{\rm V}}$ 
and $\chi_{\rm A} \equiv {\d \mu_{\rm A}}/{\d n_{\rm A}}$.

In particular, for the homogeneous magnetic field and global rotation that are aligned with 
each other, we can take ${\bm B} = B \hat {\bm x}$ and ${\bm \omega} = \omega \hat {\bm x}$ 
without loss of generality, and then the system is effectively reduced to 1+1 dimensions. 
In this case, we arrive at the terms that take exactly the same form as the chiral terms in 
Eqs.~(\ref{EFT}), where
\begin{gather}
\alpha_1 = - \frac{B}{2\pi^2 \chi_{\rm A}}\,, \quad  
\alpha_2 = \alpha_3 = - \frac{\omega}{\pi^2 \chi_{\rm V} \chi_{\rm A}}\,, \quad
\\
\beta_1 = - \frac{B}{2\pi^2 \chi_{\rm V}}\,, \quad 
\beta_2 = - \frac{\omega}{\pi^2 \chi_{\rm V}^2}\,, \quad 
\beta_3 = - \frac{\omega}{\pi^2 \chi_{\rm A}^2}\,.
\end{gather}
In this example, the chiral terms in Eqs.~(\ref{EFT}) originate from the relativistic quantum 
effects related to the chirality of elementary particles. Note here that the diffusion term 
${\bm j} = - D{\bm \nabla} n$ is higher order in derivatives compared with the CME and CVE 
if we take $B = O(\d^0)$ and $\omega = O(\d^0)$. 

In the context of these chiral transport phenomena in both high-energy physics and condensed 
matter physics, however, effects of reactions have not been taken into account. Inclusion of 
reaction terms makes the right-hand sides of Eqs.~(\ref{continuity}) nonvanishing, which may 
exhibit new and rich physical phenomena. As we will see, the interplay between chiral effects and 
reactions in Eqs.~(\ref{EFT}) gives rise to a new type of instability and helical pattern formation. 
Note again that Eqs.~(\ref{EFT}), being an effective theory based on symmetries and systematic 
derivative expansion, are not limited to this particular realization due to the CME or CVE, and 
can be relevant to generic systems involving chiral charges.

\section{Linear stability analysis}
We now study the linear stability of the system described by the effective theory (\ref{EFT}).
We assume that the system has a spatially homogeneous stable steady state 
$(n_{\rm V}$, $n_{\rm A}) = (\bar n_{\rm V}$, $\bar n_{\rm A})$, and so
$f(\bar n_{\rm V}, \bar n_{\rm A}) = g(\bar n_{\rm V}, \bar n_{\rm A}) = 0$. 
We consider sufficiently small perturbations around this steady state,
\beq
n_{\rm V} = \bar n_{\rm V} + \delta n_{\rm V}, \quad 
n_{\rm A} = \bar n_{\rm A} + \delta n_{\rm A},
\eeq
so that the reaction terms can be expanded to linear order in 
$\delta n_{\rm V}$ and $\delta n_{\rm A}$ as
\beq
f(n_{\rm V}, n_{\rm A}) = f_{\rm V} \delta n_{\rm V} + f_{\rm A} \delta n_{\rm A},  \quad
g(n_{\rm V}, n_{\rm A}) = g_{\rm V} \delta n_{\rm V} + g_{\rm A} \delta n_{\rm A},
\eeq
where 
$f_{\rm V} \equiv \left. {\d f}/{\d n_{\rm V}} \right|_{\rm ss}$, 
$f_{\rm A} \equiv \left. {\d f}/{\d n_{\rm A}} \right|_{\rm ss}$, 
$g_{\rm V} \equiv \left. {\d g}/{\d n_{\rm V}} \right|_{\rm ss}$, 
$g_{\rm A} \equiv \left. {\d g}/{\d n_{\rm A}} \right|_{\rm ss}$ 
are some constants that depend on the details of reactions (here ``ss" stands for the steady state).
From the consideration of parity symmetry, $f_{\rm V}$ and $g_{\rm A}$ are scalar while 
$f_{\rm A}$ and $g_{\rm V}$ are pseudoscalar. This is possible, e.g., if 
$f_{\rm A} \propto \bar n_{\rm A}$ and $g_{\rm V} \propto \bar n_{\rm A}$, and they can be 
nonzero when the steady state breaks parity symmetry by a nonzero $\bar n_{\rm A}$. 
Then, we have the linearized equations for Eqs.~(\ref{EFT}),
\begin{subequations}
\begin{align}
\d_t \delta n_{\rm V} &= \alpha_{\rm V} \d_x \delta n_{\rm V} + \alpha_{\rm A} \d_x \delta n_{\rm A}
+ f_{\rm V} \delta n_{\rm V} + f_{\rm A} \delta n_{\rm A}, \\
\d_t \delta n_{\rm A} &= \beta_{\rm V} \d_x \delta n_{\rm V} + \beta_{\rm A} \d_x \delta n_{\rm A}
+ g_{\rm V} \delta n_{\rm V} + g_{\rm A} \delta n_{\rm A},
\end{align}
\end{subequations}
where $\alpha_{\rm V} \equiv \alpha_2 \bar n_{\rm A}$, 
$\alpha_{\rm A} \equiv \alpha_1 + \alpha_3 \bar n_{\rm V}$, 
$\beta_{\rm V} \equiv \beta_1 + \beta_2 \bar n_{\rm V}$, 
$\beta_{\rm A} \equiv \beta_3 \bar n_{\rm A}$.
(In the presence of parity-breaking background fields, their contributions can be absorbed 
into $\alpha_{\rm V}$ and $\beta_{\rm A}$ accordingly.)
We assume $\alpha_{\rm A} \beta_{\rm V} > 0$, which ensures that the system is stable in the 
absence of reactions.

Let us take the temporally and spatially dependent (or spatially independent) perturbation 
of the form,
\beq
\delta n_{\rm V} = \epsilon_{\rm V} e^{\lambda t + iqx}, \quad 
\delta n_{\rm A} = \epsilon_{\rm A} e^{\lambda t + iqx}.
\eeq
We then get the matrix equation, $(M_{i j} + iq N_{ij} - \lambda I_{ij}) x^j =0$, where
\beq
M \equiv 
\left(
\begin{tabular}{cc}
$f_{\rm V}$ & $f_{\rm A}$ \\
$g_{\rm V}$ & $g_{\rm A}$
\end{tabular}
\right)\,,
\quad 
N \equiv 
\left(
\begin{tabular}{cc}
$\alpha_{\rm V}$ & $\alpha_{\rm A}$ \\
$\beta_{\rm V}$ & $\beta_{\rm A}$
\end{tabular}
\right)\,,
\quad
{\bm x} \equiv \left(
\begin{tabular}{c}
$\epsilon_{\rm V}$ \\
$\epsilon_{\rm A}$
\end{tabular}
\right)\,,
\eeq
and $I$ is the unit matrix.
In order for it to have a nontrivial solution, we must have $\det (M + iq N - \lambda I)= 0$, or
\begin{gather}
\label{proper}
\lambda^2 - [({\rm tr}M) + i q ({\rm tr}N)] \lambda + (\det M - q^2 \det N)
+ iq (\alpha_{\rm V} g_{\rm A} + \beta_{\rm A} f_{\rm V} 
- \alpha_{\rm A} g_{\rm V} - \beta_{\rm V} f_{\rm A}) = 0,
\end{gather}
where ${\rm tr}M = f_{\rm V} + g_{\rm A}$, ${\rm tr}N = \alpha_{\rm V} + \beta_{\rm A}$, 
$\det M = f_{\rm V} g_{\rm A} - f_{\rm A} g_{\rm V}$, and $\det N = \alpha_{\rm V} \beta_{\rm A} - \alpha_{\rm A} \beta_{\rm V}$.

The stability conditions against the spatially homogeneous perturbations can be found from 
Eq.~(\ref{proper}) with $q=0$ as
\beq
\label{cond_homo}
{\rm tr}M < 0, \quad \det M > 0.
\eeq

We now look for the condition that the steady state is unstable to a spatially inhomogeneous 
perturbation. This amounts to the condition that there must exist some real $q$, such that the 
real part of the solution to Eq.~(\ref{proper}) is positive, ${\rm Re} \lambda > 0$.
This eventually reduces to the inequality,
\beq
\label{cond_q}
\left[H(\alpha, \beta, f, g) - \alpha_{\rm A} \beta_{\rm V} ({\rm tr} M)^2  \right] q^2
> ({\rm tr} M)^2 \det M,
\eeq
where
\beq
H(\alpha, \beta, f, g) \equiv 
[\beta_{\rm V} f_{\rm A} + \alpha_{\rm A} g_{\rm V} + f_{\rm V}(\alpha_{\rm V} - \beta_{\rm A})]
[\beta_{\rm V} f_{\rm A} + \alpha_{\rm A} g_{\rm V} - g_{\rm A}(\alpha_{\rm V} - \beta_{\rm A})]\,. 
\eeq
Since the right-hand side of Eq.~(\ref{cond_q}) is some positive constant from 
Eqs.~(\ref{cond_homo}), the necessary and sufficient condition for the existence of such $q$ is
\beq
\label{cond_inhomo}
H(\alpha, \beta, f, g) > \alpha_{\rm A} \beta_{\rm V} (f_{\rm V} + g_{\rm A})^2.
\eeq
Apparently, in order for the condition (\ref{cond_inhomo}) to be satisfied, nonzero chiral terms 
are necessary. Also, the steady state must explicitly break parity symmetry (otherwise 
$f_{\rm A} = g_{\rm V} = 0$). Hence, we call this instability the ``chirality-driven instability."%
\footnote{Although the chirality-driven instability may look similar to the so-called chiral plasma instability 
in charged relativistic chiral matter \cite{Akamatsu:2013pjd}, the former is different from the latter in that it emerges 
only in the presence of reactions.}

It should be remarked that, in the original Turing mechanism \cite{Turing}, diffusion drives the 
instability and pattern formation in open systems. On the other hand, in our case, the interplay 
between chiral effects and reactions leads to the instability (and furthermore, a helical pattern 
formation, as we shall see below) {\it even without} diffusion.

For $\alpha_{\rm V} = \beta_{\rm A}$ and $\alpha_{\rm A} = \beta_{\rm V}$,%
\footnote{For the particular realization of the chiral terms due to the CME and/or CVE above, 
these two conditions are equivalent to just $\chi_{\rm V} = \chi_{\rm A}$.}
in particular, the condition (\ref{cond_inhomo}) is simplified as
\beq
\alpha_{\rm A} \neq 0, \quad |f_{\rm A} + g_{\rm V}|  > |f_{\rm V} + g_{\rm A}|\,.
\eeq
In this case, it is convenient to move to the chiral basis in terms of $n_{\rm R}$ and $n_{\rm L}$,
where the linearized equations of Eqs.~(\ref{EFT}) read
\begin{subequations}
\label{chiral}
\begin{align}
\d_t \delta n_{\rm R} &= \alpha_{\rm R} \d_x \delta n_{\rm R} + F_{\rm R} \delta n_{\rm R} + F_{\rm L} \delta n_{\rm L}, 
\\
\d_t \delta n_{\rm L} &= \alpha_{\rm L} \d_x \delta n_{\rm L} + G_{\rm R} \delta n_{\rm R} + G_{\rm L} \delta n_{\rm L},
\end{align}
\end{subequations}
where 
$\alpha_{\rm R} \equiv \alpha_{\rm V} + \alpha_{\rm A}$, 
$\alpha_{\rm L} \equiv \alpha_{\rm V} - \alpha_{\rm A}$, 
$F_{\rm R} \equiv (f_{\rm V} + f_{\rm A} + g_{\rm V} + g_{\rm A})/2$, 
$F_{\rm L} \equiv (f_{\rm V} - f_{\rm A} + g_{\rm V} - g_{\rm A})/2$, 
$G_{\rm R} \equiv (f_{\rm V} + f_{\rm A} - g_{\rm V} - g_{\rm A})/2$, and 
$G_{\rm L} \equiv (f_{\rm V} - f_{\rm A} - g_{\rm V} + g_{\rm A})/2$.
Then, the stability conditions against spatially homogenous perturbations in 
Eqs.~(\ref{cond_homo}) become
\beq
\label{cond_homo2}
F_{\rm R} + G_{\rm L}<0, \quad F_{\rm R} G_{\rm L} - F_{\rm L} G_{\rm R}>0.
\eeq
Also, the condition for the chirality-driven instability in Eq.~(\ref{cond_inhomo}) is rewritten as
\beq
\label{cond_inhomo2}
\alpha_{\rm R} \neq \alpha_{\rm L}, \quad  F_{\rm R} G_{\rm L} < 0\,,
\eeq
which suggests that one of $F_{\rm R}$ or $G_{\rm L}$ is positive while the other is negative. 
The positive one may be called the {\it activator}, and the negative one the {\it inhibitor}. This 
has a somewhat similar structure to the Turing's activator-inhibitor model \cite{Turing, Murray}, 
although the underlying mechanism leading to the instability is different.

\section{Helical pattern formation}
As a demonstration of the chirality-driven instability and helical pattern formation, 
we consider the following toy model for chiral charges, $n_{\rm R}(x,t)$ and $n_{\rm L}(x,t)$:
\begin{subequations}
\label{model}
\begin{align}
\d_t n_{\rm R} &= \alpha_{\rm R} \d_x n_{\rm R}  + F_{\rm R} n_{\rm R} + F_{\rm L} n_{\rm L} + \gamma n_{\rm R}^3, 
\\
\d_t n_{\rm L} &= \alpha_{\rm L} \d_x n_{\rm L} + G_{\rm R} n_{\rm R} + G_{\rm L} n_{\rm L},
\end{align}
\end{subequations}
where all the variables and coefficients are made dimensionless by certain rescaling 
(and, for simplicity of notation, we use the same variables and coefficients as before.) 
This model may be seen as describing fluctuation of chiral charges around the the 
parity-breaking background field (e.g., background axial charge $n_{\rm A}^0 \neq 0$) for 
Eqs.~(\ref{chiral}). Here, the nonlinear $\gamma$ term is also added to stabilize the system.

\begin{figure}[t]
\begin{center}
\includegraphics[width=7.5cm]{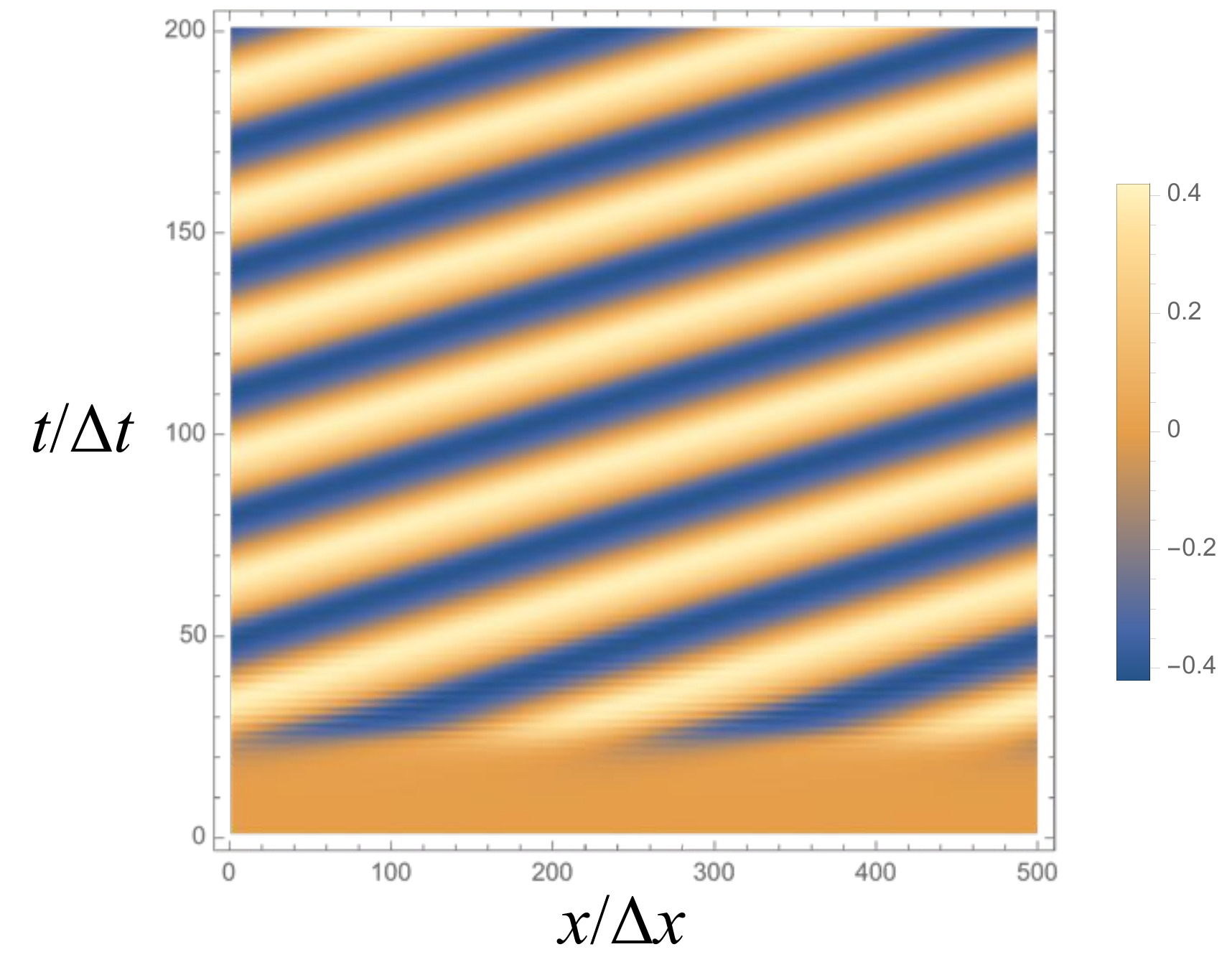}
\includegraphics[width=7.5cm]{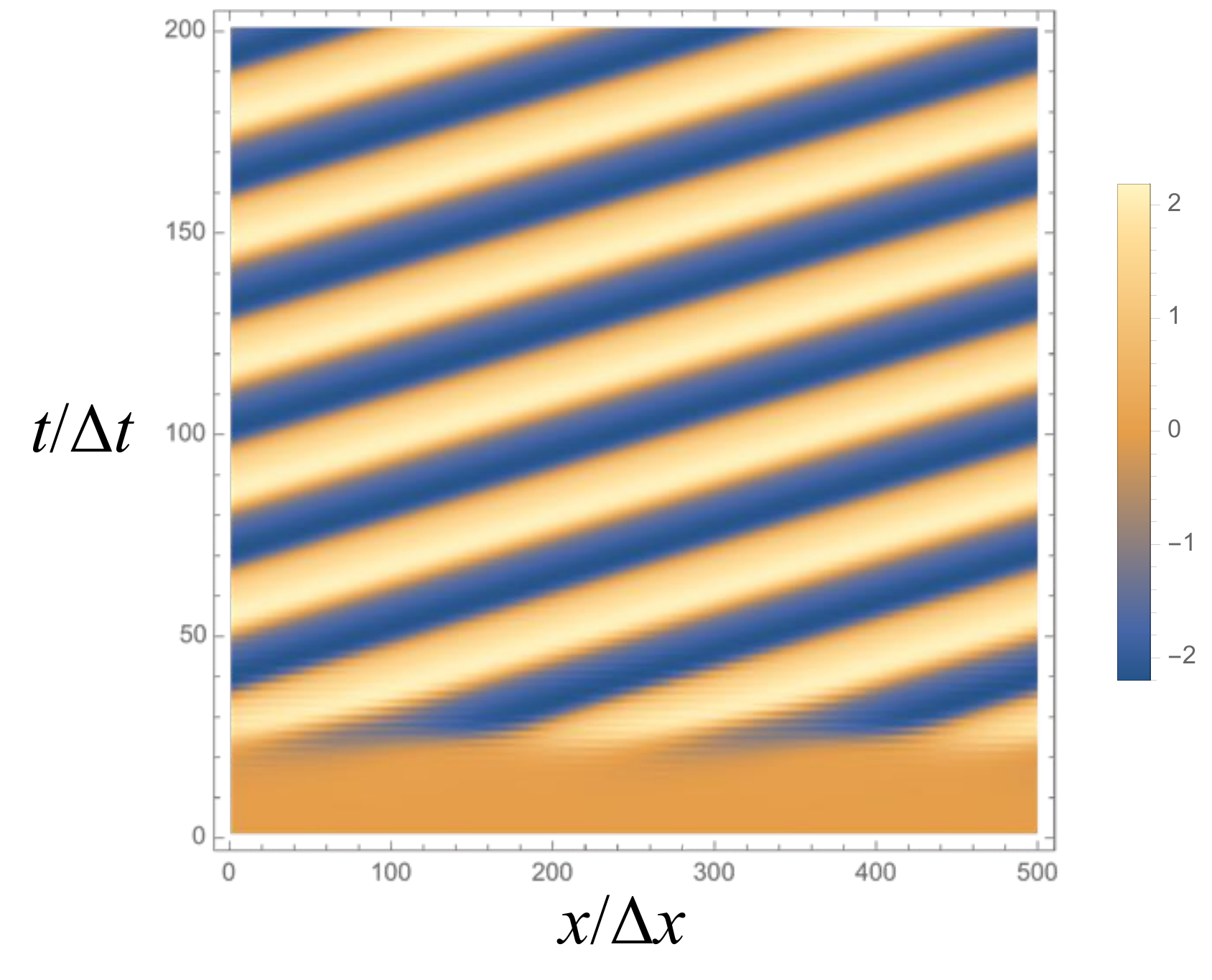}
\end{center}
\vspace{-0.7cm}
\caption{Plots of $n_{\rm L}$ (left) and $n_{\rm R}$ (right) for the toy model (\ref{model}) in the $(x,t)$ plane.}
\label{fig:helical}
\end{figure}

As an example, we take 
$\alpha_{\rm R} = 0.5$, $\alpha_{\rm L} = 0.1$, $F_{\rm R} = 0.5$, $F_{\rm L} = -1.5$, 
$G_{\rm R} = 1$, $G_{\rm L} = -2$, and $\gamma = -0.1$, such that the conditions 
(\ref{cond_homo2}) and (\ref{cond_inhomo2}) are satisfied. As an initial condition, we take 
a perturbation $n_{\rm R}(x,0) = \epsilon r_{\rm R}(x)$ and $n_{\rm L}(x,0) = \epsilon r_{\rm L}(x)$, 
where $r_{\rm R,L}(x)$ are uniform random functions in the interval $[-1, 1]$ and $\epsilon=0.01$.  
In order to compute the time evolution of the initial perturbation, we perform the numerical 
simulation for 200 units of time with time step size $\Delta t=0.002$ and for 500 grid points 
with spatial grid size $\Delta x=0.002$. Then, we find $n_{\rm R, L}(x,t)$ at each spatial grid point 
$x=k \Delta x$ ($k=1,2,\cdots,500$) and at each time step $t=\ell \Delta t$ ($\ell=1,2,\cdots,200$).

The numerical result in the $(x,t)$ plane is shown in Fig.~\ref{fig:helical}. This shows that the 
initial perturbation evolves into a propagating helical pattern in the $(x,t)$ plane%
\footnote{So far, we have focused on the $(x,t)$ coordinates in 1+1 dimensions. We can 
immediately extend this result to 3+1 dimensions when the system has inertia in the other 
($y$ and/or $z$) direction. Then, this propagating state in 1+1 dimensions corresponds to 
a helical pattern in 3+1 dimensions.}---a feature that cannot be seen in the usual Turing 
pattern in 1+1 dimensions. This emergent macroscopic helical structure spontaneously breaks 
parity and both temporal and spatial translational symmetries (although the initial state does not). 
In particular, this result shows that microscopic chirality can generate a macroscopic helical 
structure via nonequilibrium processes even in the absence of diffusion. 

This symmetry breaking pattern is reminiscent of the so-called ``chiral soliton lattice" (CSL) 
that breaks parity and spatial translational symmetries. The CSL is realized as the ground state 
of various physical systems with chirality, such as cholesteric liquid crystals \cite{DeGennes}, 
chiral magnets \cite{Dzyaloshinskii, CM}, and quantum chromodynamics at finite density in a 
magnetic field \cite{Brauner:2016pko} and/or under a rotation \cite{Huang:2017pqe}. Compared 
with the CSL in these systems, the parity- and translation-violating structure emerges via 
nonequilibrium processes in the present case. In passing, we also note that the resulting state 
here may be regarded as a nonequilibrium realization of the so-called time crystal \cite{Shapere:2012nq, Richerme}.

\section{Discussion and outlook}
In this paper, we have demonstrated a new mechanism of chirality-driven instability and helical 
pattern formation due to the interplay between the chiral effects and reactions in open systems. 
Unlike the original Turing's diffusion-driven instability, diffusion is irrelevant in our mechanism. 
Our mechanism, being based on the effective theory, is relevant to generic systems with chiral 
charges, as long as the conditions (\ref{cond_homo}) and (\ref{cond_inhomo}) are satisfied. 
In particular, it is applicable to chiral charges of elementary particles, where the emergence of 
macroscopic helical patterns is a consequence of topological currents (CME and/or CVE) 
associated with their chirality.

There are several future directions in which one can extend our analysis.
(i) Effects of diffusion can be included as the higher-order correction to see how our mechanism 
is quantitatively modified. This study would clarify the competition or interplay between the
chirality-driven instability and diffusion-driven instability.
(ii) It is straightforward to generalize our effective theory (\ref{EFT}) to 3+1 dimensions to study 
the helical pattern formation. One such direction is to solve Eqs.~(\ref{continuity_chiral}) for 
inhomogeneous magnetic fields and/or rotation in the presence of reactions. 
(iii) Since chirality is defined only in odd spatial dimensions, our mechanism of the 
chirality-driven instability is not directly applicable in two spatial dimensions. Still, one can ask if 
and how the topological currents in 2+1 dimensions, such as the quantum Hall effect, can lead 
to pattern formation without diffusion.

Finally but not least, it is an important question to understand whether and how our mechanism 
can be realized in actual physical systems. Concerning the chirality of elementary particles, it is 
typically considered that effects of parity violation by the weak interaction are too small to affect 
macroscopic helical structures \cite{Bonner2000, Avalos2000}. Even if so for each microscopic 
weak process, this is not necessarily the case in astrophysical systems. In fact, it has been 
recently argued that a core-collapse supernova is the system with the macroscopically largest 
parity violation in the Universe, where nonequilibrium electron capture reactions, 
${\rm p} + {\rm e}_{\rm L}^- \rightarrow {\rm n} + \nu^{\rm e}_{\rm L}$, involving only left-handed 
electrons and neutrinos, produce large chirality asymmetries of leptons, and consequently, 
a strong helical magnetic field (which is equivalent to circularly polarized light), 
and helical fluid motion \cite{Yamamoto:2015gzz}. Moreover, the length scale of magnetic fields 
and fluid motion there can be amplified to macroscopic scales by the inverse cascade of the 
chiral turbulence \cite{Masada:2018swb}. Since the supernova is an open system with such 
large parity violation, our mechanism of the helical pattern formation may potentially be realized. 
This is just one possible example, and it would be interesting to investigate the relevance in 
other systems as well.

\section*{Acknowledgement}
The author thanks Kouichi Asakura, Katsuya Inoue, Jun-ichiro Kishine, and Shigeru Kondo 
for useful conversations. 
This work was supported by JSPS KAKENHI Grant No.~16K17703, MEXT-Supported 
Program for the Strategic Research Foundation at Private Universities, ``Topological Science" 
(Grant No.~S1511006), and JSPS Core-to-Core Program, A. Advanced Research Networks.


\begin{thebibliography}{99}

  \bibitem{Bonner2000}
  W.~A.~Bonner, Chirality {\bf 12}, 114 (2000).
  
  \bibitem{Avalos2000}
  M.~Avalos, R.~Babiano, P.~Cintas, J.~L.~Jimenez, and J.~C.~Palacios, 
  Tetrahedron: Asymmetry {\bf 11}, 2845 (2000).

  \bibitem{Turing}
  A.~M.~Turing, Phils.\ Trans.\ R.\ Soc. London Ser. B, {\bf 273}, 37 (1952).
  
  \bibitem{Murray}
  J.~D.~Murray, {\it Mathematical Biology II: Spatial Models and Biomedical Applications}, 
  3rd ed. (Springer-Verlag, New York, 2003).

  \bibitem{Vilenkin:1980fu}
  A.~Vilenkin,
  Phys.\ Rev.\ D {\bf 22}, 3080 (1980).
  
  \bibitem{Nielsen:1983rb} 
  H.~B.~Nielsen and M.~Ninomiya,
  Phys.\ Lett.\ B {\bf 130}, 389 (1983).  

  \bibitem{Fukushima:2008xe}
  K.~Fukushima, D.~E.~Kharzeev, and H.~J.~Warringa,
  Phys.\ Rev.\ D {\bf 78}, 074033 (2008).  
  
  \bibitem{Vilenkin:1979ui} 
  A.~Vilenkin,
  Phys.\ Rev.\ D {\bf 20}, 1807 (1979).
  
  \bibitem{Son:2009tf}
  D.~T.~Son and P.~Sur\'owka,
  Phys.\ Rev.\ Lett.\  {\bf 103}, 191601 (2009).
  
  \bibitem{Landsteiner:2011cp} 
  K.~Landsteiner, E.~Megias, and F.~Pena-Benitez,
  Phys.\ Rev.\ Lett.\  {\bf 107}, 021601 (2011).
      
  \bibitem{Son:2012wh}
  D.~T.~Son and N.~Yamamoto,
  Phys.\ Rev.\ Lett.\  {\bf 109}, 181602 (2012).

  \bibitem{Akamatsu:2013pjd} 
  Y.~Akamatsu and N.~Yamamoto,
  Phys.\ Rev.\ Lett.\  {\bf 111}, 052002 (2013).
  
  \bibitem{DeGennes}
  P.~G.~de~Gennes, Solid State Commun. {\bf 6}, 163 (1968).
  
  \bibitem{Dzyaloshinskii} 
  I.~E.~Dzyaloshinskii, Zh.\ Eksp.\ Teor.\ Fiz.\ {\bf 46}, 1420 (1964) [Sov.\ Phys.\ JETP {\bf 19}, 960 (1964)].
  
  \bibitem{CM}
  Y.~Togawa, T.~Koyama, K.~Takayanagi, S.~Mori, Y.~Kousaka, J.~Akimitsu, S.~Nishihara, K.~Inoue, A.~S.~Ovchinnikov, and J.~Kishine, 
  Phys.\ Rev.\ Lett.\ {\bf 108}, 107202 (2012).
  
  \bibitem{Brauner:2016pko}
  T.~Brauner and N.~Yamamoto,
  JHEP {\bf 1704}, 132 (2017).
  
  \bibitem{Huang:2017pqe} 
  X.~G.~Huang, K.~Nishimura, and N.~Yamamoto,
  JHEP {\bf 1802}, 069 (2018).
    
  \bibitem{Shapere:2012nq} 
  A.~Shapere and F.~Wilczek,
  Phys.\ Rev.\ Lett.\  {\bf 109}, 160402 (2012);
  F.~Wilczek,
  Phys.\ Rev.\ Lett.\  {\bf 109}, 160401 (2012).
  
  \bibitem{Richerme}
  P.~Richerme, 
  Physics {\bf 10}, 5 (2017).
  
  \bibitem{Yamamoto:2015gzz} 
  N.~Yamamoto,
  Phys.\ Rev.\ D {\bf 93}, 065017 (2016).
  
  \bibitem{Masada:2018swb} 
  Y.~Masada, K.~Kotake, T.~Takiwaki, and N.~Yamamoto,
  arXiv:1805.10419 [astro-ph.HE].

\end{thebibliography}
\end{document}